\documentclass[doublecol]{epl2} 

\title{Thermodynamically stable noncomposite vortices in mesoscopic two-gap 
superconductors}
\shorttitle{Thermodynamically stable noncomposite vortices in mesoscopic two-gap 
superconductors} 

\author{L.F. Chibotaru, V.H. Dao \and A. Ceulemans}
\shortauthor{L.F. Chibotaru, V.H. Dao and A. Ceulemans}

\institute{Division of Quantum and Physical Chemistry, and \\
INPAC - Institute for Nanoscale Physics and Chemistry, \\ 
University of Leuven,  Celestijnenlaan 200F, B-3001 Leuven, Belgium 
}
\pacs{74.78.Na}{Mesoscopic and nanoscale systems}
\pacs{74.20.De}{Phenomenological theories (two-fluid, Ginzburg-Landau, etc.)}
\pacs{74.25.Ha}{Magnetic properties}

\abstract{In mesoscopic two-gap superconductors with sizes of the order of the
coherence length  noncomposite vortices are found to be thermodynamically stable
in a large domain of the $T - H$ phase diagram.
In these phases the vortex cores of one condensate are spatially separated from 
the other condensate ones, and their respective distributions can adopt distinct symmetries. 
The appearance of these vortex phases is caused by a
non-negligible effect of the boundary of the sample on the superconducting order parameter and 
represents therefore a genuine mesoscopic effect. 
For low values of interband Josephson coupling vortex patterns with $L_1 \neq L_2$ can arise in
addition to the phases with $L_1 =L_2$, where
$L_1$ and $L_2$ are total vorticities in the two condensates.
The calculations show that
noncomposite vortices could be observed in thin mesoscopic 
samples of MgB$_{2}$.}

\begin{document}

\maketitle

Two-gap superconductivity 
started to attract much attention in
connection with the discovery of the MgB$_{2}$ superconductor, for which a 
clear-cut
evidence for the existence of two gaps was obtained \cite{mgb2}. 
Superconductors of this kind show new
qualitative effects with respect to conventional ones. First, the presence of
two or more electronic bands at the Fermi level always enhances the
superconductivity as compared to the effect from any individual 
band. Second, a new phenomenon, the fractionalization of the
magnetic flux associated to individual vortices in massive two-gap
superconductors is predicted  \cite{babaev,tanaka}. The condition for
this fractionalization is the inequality of the winding numbers of the 
vortices in the
two condensates ($L_1 \neq L_2$) having a common vortex core (composite
vortices). Although these vortex phases have finite energies per unit length
they never correspond to the ground state, i.e. are thermodynamically unstable
\cite{babaev,tanaka}. On the other hand the energy per unit length of a
vortex configuration where the vortices in each of the two bands
are spatially separated (noncomposite or deconfined vortices) was found to
be divergent \cite{babaev}. These results are in line with the fact that
only composite vortices with $L_1 = L_2 =1$, i.e. usual Abrikosov 
vortices, have been observed in massive two-gap superconductors to date. 

Fractionalization of the vortex magnetic flux can also be observed in layered superconductors 
through thermal fluctuations \cite{decol}. By the action of the interlayer Josephson coupling and 
the magnetic field, pancake vortices (which individually live inside one layer) align themselves 
into vortex stacks which thread across the layers. Since each pancake vortex carries
only $\approx \Phi_0/N_l$ (where $N_l$ is the number of layers) the dissociation of a 
vortex stack results in a net fraction of the flux quantum. This mechanism is however different
 from the one in a multigap superconductor where a fractional flux vortex lives within one condensate
but threads through the entire sample thickness. 

In mesoscopic superconductors, the geometry of the confinement of the
superconducting condensate influences essentially their properties because
the coherence length, $\xi (T)$, and the penetration length, $\lambda (T)$,
become of the order of the samples size (for a review, see,
e.g., \cite{moshchalkov,chibotaru}). In
particular, new vortex patterns arise as function of boundary geometry
leading to cusp-like normal-superconducting phase boundary, $T_c (H)$
\cite{moshchalkov1,chibotaru}, and to the nucleation of giant vortices in
disks \cite{giant,peeters} and of antivortices in 
regular polygons \cite{chibotaru,meso}. 
Another
important mesoscopic effect is that the vortices nucleate in patterns which
are
quite different from an Abrikosov lattice and approach the latter by a
series of phase transitions when temperature is lowered from $T_c (H)$
\cite{peeters,chibotaru1}. The symmetry of nucleated vortex patterns is
always consistent with the geometry of the boundary of the sample
\cite{chibotaru}, while the region of their stability against Abrikosov type
vortices increases with diminishing size of the sample, persisting down
to $T=0$ when the samples reach some critical size \cite{chibotaru1}.
  
In this Letter we investigate the vortex patterns in thin mesoscopic
two-gap superconductors. Contrary to what was found in the case of massive 
superconductors, we show that noncomposite vortices can arise as 
thermodynamically stable phases in mesoscopic samples of small enough size. 
We find
the reasons for their stabilization and discuss
the possibility of their experimental observation.

Consider a superconducting sample of size $R$ and thickness $d$ in a
perpendicular uniform magnetic field $H$.
For small ($R\sim\xi$) and thin ($d\ll\xi < \lambda$) samples one can neglect the 
variation of the order
parameter across thickness and the distortion of the magnetic 
field induced by screening and vortex currents\footnote{  
For a thin superconductor with 
$\Psi({\bf r})=\sqrt{\frac{|\alpha|}{\beta}}f({\bf r}) e^{i\chi({\bf r})}$, the variation of the vector potential
due to the supercurrent $\delta \!{\bf A}({\bf r}_0) \approx  - \frac{d}{4\pi \lambda^{2}} 
\int (\!\frac{\Phi_0}{2\pi}\!\nabla\! \chi\! + \!{\bf A}\!)\!f^2\!({\bf r})/|{\bf r}_0-{\bf r}| d\!S
\sim - \frac{d R}{2 \lambda^{2}} 
(\!\frac{\Phi_0}{2\pi}\!\nabla\! \chi\! + \!{\bf A}\!)\!f^2\!({\bf r}_0)$ 
 can be neglected if the thickness $d$ is small enough. The same argument can be 
 generalized to a two-gap superconductor. } \cite{chibotaru,moshchalkov1,peeters}. The two components of the order parameter
$\Psi_{1,2}$ \cite{comment} are found from the minimization of the 
following 2D Ginzburg-Landau functional \cite{dao}:
\begin{eqnarray}
\Delta F = && \int \Biggl[ \sum_{n=1}^{2} \biggl(  
\frac{1}{2m_n}\bigl|\bigl( -i\hbar\nabla
-\frac{2e}{c}{\mathbf A}\bigr) \Psi_n\bigr|^2 \bigr)+\alpha_n |\Psi_n |^2
\nonumber\\
&& + \frac{1}{2}\beta_n |\Psi_n |^4 \biggr)
-\gamma \bigl( \Psi_1^*\Psi_2+\Psi_2^*\Psi_1 \bigr) \Biggr] dS ,
\label{functGL}
\end{eqnarray}
where $\mathbf A$ is the vector potential of the applied field, $\alpha_1=-a_1
t$ and $\alpha_2=\alpha_{20}-a_2 t$ are the condensation energy parameters
for the active and the passive band, respectively, $t\approx 1-T/T_1$, and 
$T_1$ 
is the critical temperature corresponding to the active band \cite{dao}. 
At the normal/superconducting phase transition, minimizing (\ref{functGL}) (without the terms $\sim
|\Psi_n |^4$) results in two linear equations describing the nucleation of
superconductivity and two boundary conditions:
\begin{eqnarray}
&&\bigl[ \alpha_n +\frac{1}{2m_n} \bigl( -i\hbar\nabla-\frac{2e}{c}{\mathbf A}
\bigr)^2 \bigr] \Psi_n 
-\gamma \Psi_{n'}= 0 , \nonumber\\
&& \bigl( -i\hbar\nabla-\frac{2e}{c}{\mathbf A}\bigr) {\Psi_n}_
{|_{\rm{n.b.}}}=0,
\label{nucleation}
\end{eqnarray}
where $n,n'$=1,2 and 2,1. The notation $|_{\rm{n.b.}}$ means the expression is projected on the unit vector normal to the boundary. It is straightforward to show that the nucleation solution
of these equations has the form $\Psi_1 \sim \Psi_2 \sim \phi_N$,
where $\phi_N$ is the solution of the eigenvalue equation:
\begin{eqnarray}
&&\bigl( -i\hbar\nabla-\frac{2e}{c}{\mathbf A}\bigr)^2 \phi_i = \lambda_i
\phi_i , \nonumber\\
&& \bigl( -i\hbar\nabla-\frac{2e}{c}{\mathbf A}\bigr) {\phi_i}_
{|_{\rm{n.b.}}}=0,
\label{eigen}
\end{eqnarray}
corresponding to $\lambda_N (H)$ which is the lowest of the eigenvalues $\lambda_i (H)$. This eigenvalue
determines the nucleation phase boundary $T_c (H)$ \cite{dao} and  
the ratio of the amplitudes of the two components of the order parameter,
$\Psi_2 /\Psi_1 =\gamma /(\alpha_2 +\lambda_N /2m_2 )$. Thus the distribution of
the nucleated order parameter in the two bands of a two-gap
superconductor is described by the same function. 

The eigenstates of (\ref{eigen})
are then used as the basis set for the order parameter of the complete functional (\ref{functGL}),
\begin{equation}
\Psi_1 = \sum_i u_i \phi_i ,\;\;\;\;\;\;\;\;
\Psi_2 = \sum_i v_i \phi_i ,
\label{psi}
\end{equation}
which yields $\Delta F$ as a function of the expansion coefficients $\{u_i ,
v_i \}$.

We will consider further a disk of radius $R$. It is then convenient to 
introduce new "lengths" defined by the relations: 
$-\alpha_1=\hbar^2/2m_1 \xi_1(T)^2$, $\alpha_{20}=\hbar^2/2m_1
\xi_{20}^2$, $-a_2 t =\hbar^2/2m_1\xi_{2}(T)^2$, and $\gamma
=\hbar^2/2m_1\xi_{\gamma}^2$. When $\gamma =0$ and $\alpha_2 >0$ the
superconductivity nucleates in the active band, for which the coherence
length is $\xi_1$. The order parameter
corresponds to one single giant vortex in the center of the disk
\cite{giant} whose winding number increases with the applied field
following the cusps on the nucleation phase boundary line (Fig. 1a).
At lower temperatures it undergoes a broken-symmetry phase transition to
a multivortex state corresponding to Abrikosov ($L=1$) vortices forming a
regular polygon \cite{peeters}. Since $1/\xi_1^2 \sim T_c -T$, it follows 
that the temperature region where the nucleated (giant vortex) phase is 
thermodynamically stable (dark regions in Fig. 1a) scales with the size of 
the disk as $\sim R^{-2}$. 
Note that for $R \ge 6\xi_1$ new broken-symmetry phases arise, which
correspond to off-center displacements of the central vortex in the
case $L=1$ and of the maximum of $|\Psi |^2$ distribution in the case
$L=0$ \cite{comment1}.

\begin{figure}
\begin{center}
 \scalebox{.49}{%
 \includegraphics*{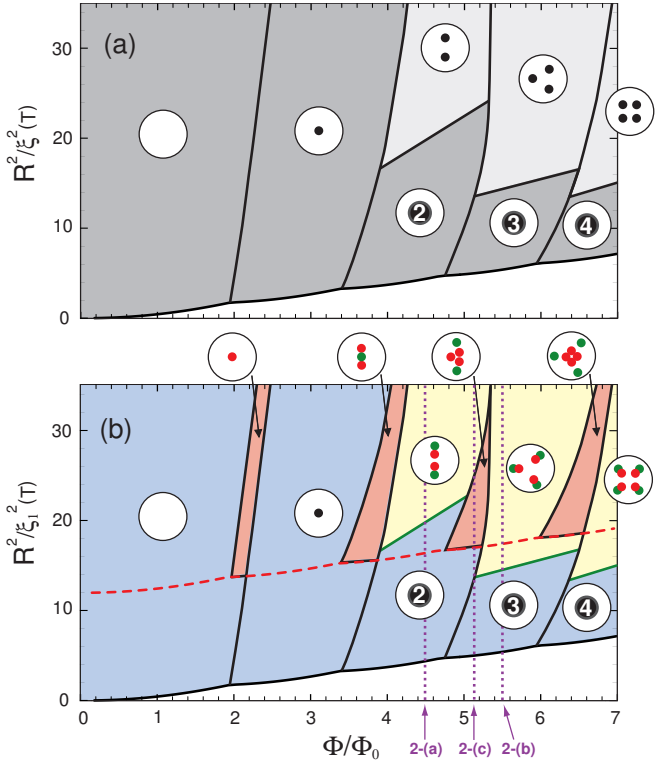}}
\end{center}
\caption{(color) Phase diagram for a thin disk of single-gap (a) and two-gap
(b) superconductors. The phase diagram (a) is parameter-free \cite{chibotaru1},
while the diagram (b) is for $a_1 /a_2 = \beta_1 /\beta_2 = m_1 /m_2 =1$,
$R/\xi_2 (0)=2\sqrt{3}$ and $R/\xi_{\gamma} (0)=0.1$. 
For each phase, the vortex structure is shown
schematically in black for composite vortices and in green (active band) and 
red (passive band) for noncomposite ones. The numbers inside the circles indicate
the vorticities of the giant vortex phases.
Vertical lines separate phases with different total vorticity $L$. Continuous
horizontal lines correspond to broken-symmetry phase transitions, which are
of giant vortex - multivortex type for $L>1$. The dashed line is the nucleation  
phase boundary for the passive band for $\gamma =0$ (see the text). The vertical dotted 
lines mark the positions of the three graphics in Fig. 2.
}
\end{figure}
In the case of nonzero interband Josephson coupling similar phase diagrams
emerge (Fig. 1b). We see again regions corresponding to giant vortex patterns (shown 
in blue in Fig. 1b), with the same vortex numbers in both condensates, and regions
with broken-symmetry vortex patterns (yellow and pink regions in Fig. 1b) consisting of
Abrikosov vortices. 
However a qualitatively new feature arises in the broken-symmetry phases, with two vortex 
patterns,
corresponding to $\Psi_1$ and $\Psi_2$, being {\em spatially distinct} for any
parameters of (\ref{functGL}). Therefore each vortex in these phases 
is a {\em noncomposite vortex}. Moreover, in contrast to single band superconductors 
(Fig. 1a), we have now two types of multivortex phases, corresponding to $L_1 =L_2$
(yellow regions in Fig. 1b) and $L_1 \neq L_2$ (pink regions in Fig. 1b). The corresponding
order parameter's density plots are shown in Fig. 2.

The temperature evolution of the vortex patterns in two condensates is strongly dependent
on the parameter $\alpha_2$, while the difference in their shape increases with diminishing 
$\gamma$. Indeed, as Fig. 2 shows with $R/\xi_{\gamma} (0)=0.1$, for small values of $\gamma$ we encounter three 
qualitatively different situations. In the first case (a) the transition to the broken-symmetry
phase leads to a strong separation of the multivortex patterns, corresponding to the active
(green) and the passive (red) bands, the latter approaching the first one with further lowering
of temperature. In the second case (b) the vortex pattern in the passive band first shrinks,
resulting in its strong separation from the vortex pattern in the active band, and then 
approaches it again when the temperature is lowered. Finally, the case (c) corresponds to the
transition to the broken-symmetry phase phase with $L_1 \neq L_2$, followed by the transition 
to a broken-symmetry phase with $L_1 =L_2$ when the temperature is lowered further. 

\begin{figure}
 \scalebox{.51}{%
\includegraphics*{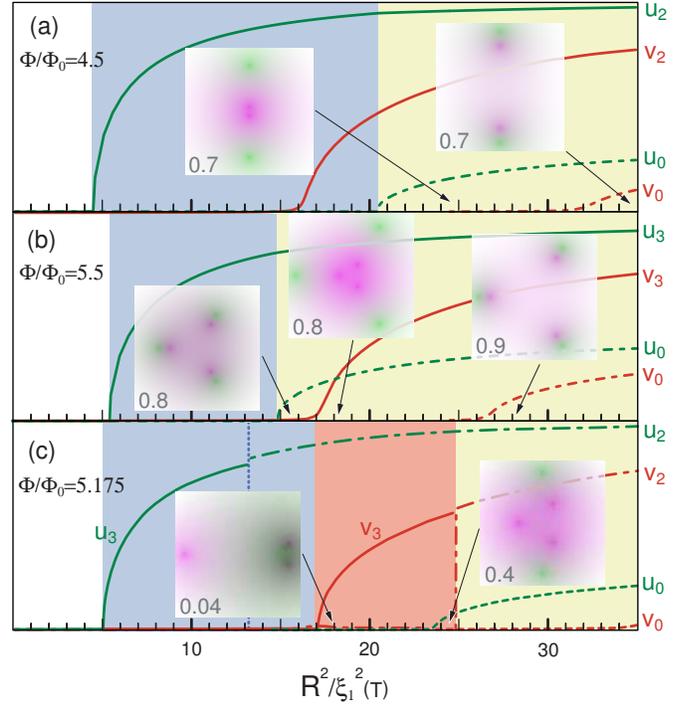}}
\caption{(color) (a) Evolution of the order parameter for three values of the applied
magnetic flux in the phase diagram of Fig. 1b. The background colours correspond to the
giant vortex phases (blue) and the multivortex phases with $L_1 =L_2$ (yellow) and 
$L_1 \neq L_2$ (pink). The expansion coefficients $u_L$ and $v_L$ correspond to the
ground state solutions of (\ref{eigen}) for each $L$ that contribute 
substantially to the expansion. The numbers in the corner of
inserted plots gives the side of the zoom on the disk in units of $R$.  
The increase of the intensity of colors corresponds to the
decrease of the density of the order parameters, so that the darkest regions
denote the position of the vortex cores.   
}
\end{figure}
The reason for the above behaviour of the order parameter 
can be elucidated by the following
simple consideration \cite{chibotaru1}. Let's suppose for simplicity that only 
one eigenstate of (\ref{eigen}) ($\sim \phi_A$) admixes to the nucleated order
parameter ($\sim \phi_N$) in the point of broken-symmetry transition,
which is actually a reasonable approximation for disks \cite{peeters} and
regular polygons \cite{chibotaru1}. When $\gamma$ is small the nucleation 
of superconductivity takes place at $-\alpha_1\approx \lambda_N /2m_1$. 
At the onset of the transition to the 
broken symmetry phase we have $\Psi_1 =u_N \phi_N + u_A \phi_A$ 
($\phi_A$ is
supposed to be of different symmetry compared to $\phi_N$, which in the
case of disk implies different rotational quantum numbers). 
If $\alpha_2 + \lambda_N /2m_2 >0$ (the sufficient condition for this is $\xi_2
(0)>\xi_{20}$), the minimization of (\ref{functGL}) gives for the
expansion coefficients of $\Psi_2$:
\begin{equation}
v_i \approx
\frac{\beta_1}{\beta_2}\frac{\xi_1^2}{\xi_{\gamma}^2}
\frac{1}{\frac{\xi_1^2}{\xi_{20}^2}-\frac{\xi_1 (0)^2}{\xi_{2}(0)^2}+
\frac{m_1}{m_2}\frac{\xi_1^2}{R^2}\epsilon_i} u_i ,
\label{induced}
\end{equation}
where $\epsilon_i= \lambda_i R^2/\hbar^2$ are dimensionless eigenvalues of (\ref{eigen}), only $\xi_1$ is temperature dependent, and the dependence on $R$ is explicit.
If $v_i$ and $u_i$ were related by the same proportionality 
coefficient, the order parameters $\Psi_1$ and $\Psi_2$ would have been
described by essentially the same function, in particular, all vortices
arising in such a phase would correspond to composite ones. Such a
situation obviously occurs in the giant vortex phase. However, as
Eq.(\ref{induced}) shows, the transition to a broken-symmetry phase makes
the coefficients of proportionality different for the admixed ($i=A$) and the
nucleated ($i=N$) components of the order parameter due to the term $\sim\epsilon_i$ 
in the denominator of (\ref{induced}). This term becomes important for
small values of $R$, i.e. in
the mesoscopic regime, and it disappears 
in the macroscopic regime $R\rightarrow\infty$. This is in line
with the conclusion mentioned in the introduction, that only composite
vortices can arise in thermodynamically stable phases of massive two-gap
superconductors.

The above analysis allows to explain the phase diagram in Fig. 1b.
For large positive $\alpha_2 + \lambda_N /2m_2$ the superconductivity in
the passive band is induced by the Josephson coupling to the active band.
Then the order parameter and the resulting vortex pattern in the passive
band follow closely that of the active band, according to Eq.
(\ref{induced}). However when this quantity is small and can
turn to negative at some temperature $T^*$ (shown by dashed line in Fig. 1b) 
then for $T<T^*$ the nucleated component $\phi_N$ in the passive band begins to grow much
faster than in the previous case due to the intrinsic superconductivity which
now exists in the passive band. Then two qualitative situations can occur. If at
$T=T^*$ the superconductor is still in the giant vortex phase, the
subsequent transition to the multivortex phase in the active band will not
induce a similar order parameter in the passive band. The latter will
remain almost $\sim\phi_N$ and consequently a much smaller
amplitude of the splitting of the giant vortex into Abrikosov vortices is
expected (Fig. 2a). If at $T=T^*$ the system is already in the multivortex state,
then the subsequent lowering of temperature will induce the shrinkage of the
vortex pattern in the passive band due to a faster stabilization of the
nucleated component in $\Psi_2$. This is exactly what is seen in Fig.
2b. In both these situations a strong spatial separation of the vortex
patterns is achieved. 
The existence of the vortex patterns with $L_1 \neq L_2$ is merely due to the fact
that for $\gamma =0$ the phase diagram for the passive band is shifted upwards 
relative to the active one 
(the nucleation of superconductivity takes place at lower temperatures), which creates
overlap regions where the two condensates have different voticities.
If $\gamma$ is smaller than the difference of free energies corresponding to
$L_1$ and $L_2$ in the passive band then vortex patterns with different vorticities 
in the condensates will be stabilized also for $\gamma \neq 0$, but will disappear
from the phase diagram when $R/ \xi_{\gamma}$ exceeds some critical value. 
The existence of the overlap regions with $L_1\neq L_2$ is possible due to the fact that 
the lines separating domains of different vorticity on the diagram on Fig. 1 are not 
vertical. As a consequence, lowering the temperature at fixed applied flux can result in 
several phase transitions with change of vorticity in the condensates, accompanied by 
jumps of the order parameter, i.e. of the coefficients $u_i$ and $v_i$ in Eq. (4) (Fig. 2c).

\begin{figure}
 \scalebox{.64}{%
 \includegraphics*{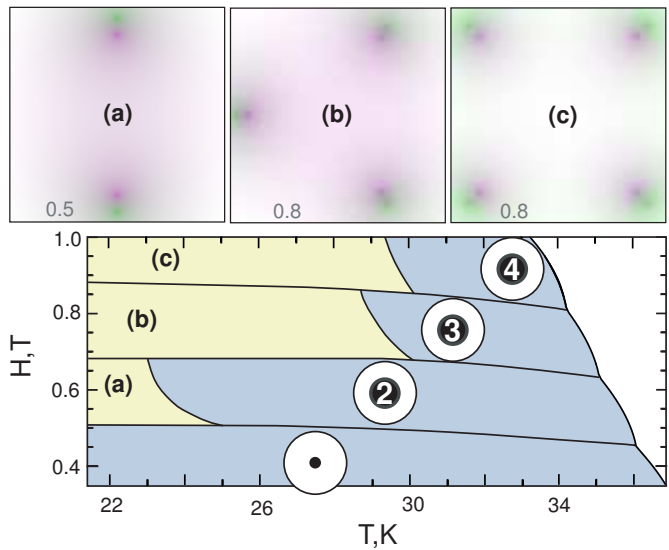}}
\caption{(color) The phase diagram for a thin disk of MgB$_2$ with the
radius R=70 nm and the parameters taken from Ref. \cite{dao,tarantini}. The upper
panels are zooms on the distribution of the order parameter in broken-symmetry
(multivortex) phases. The conventions
follow Figs. 1 and 2.
}
\end{figure}

Finally, we calculate the $T - H$ phase diagram for a thin disk of MgB$_2$ 
with the anisotropy axis $c$ perpendicular to the plane of the sample\footnote{This is the way in which thin MgB$_2$ films actually
grow on the substrate.} (Fig.3). Formation of noncomposite vortices is favored by a
weak interband coupling $\gamma$ and a small mass ratio $m_2/m_1$. From current
available data, the smallest $\gamma$ is obtained for a sample of Boron isotope Mg$^{10}$B$_2$
with a value reduced by 30\% from usual Mg$^{11}$B$_2$\cite{walte}, while
$m_{\pi}/m_{\sigma}<0.1$ has been observed in irradiated samples\cite{tarantini}
(where the introduced disorder should change effective masses but not the
coupling constants). Modifying consequently the mass ratio in the Ginzburg-Landau parameters
estimated previously for pristine MgB$_2$\cite{dao}, we use
$a_2/a_1=\beta_2/\beta_1=1.5$, $m_2/m_1=0.07$, $R/\xi_1(0)=10$,
$R/\xi_{20}=8.1$, $R/\xi_{\gamma}=6.3$ and $T_c=39$K for Fig. 3 \cite{lambda}. 
We can see several regions on the phase diagram where the two vortex
patterns are well separated in space, some of them being shown in Figs. 3 a-c.
The radius of the disk was taken $R=70$ nm but similar separations of the
vortex patterns (3 - 5 nm) were found for a wide range of radii: $R$=30 - 120 nm. 
We have observed that other
sets of parameters are not critical for the separation of the two vortex patterns
except for the ratio $m_2/m_1$: a smaller ratio leads to a higher separation.

The existence of noncomposite vortex patterns in two-gap superconductors can
be experimentally verified by combining different local probe techniques \cite{kanda,eskildsen}. 
The discussed effects are not related to the symmetry of the samples but arise due to 
non-negligible influence of the boundary of the sample on the two superconducting condensates. 
In this sense the emergence of noncomposite vortices found in the present study 
represents a true mesoscopic effect, not observable in infinite superconductors. We note in this 
connection that noncomposite vortices could also be stabilized in ultracold atomic Fermi gases 
in optical lattices or in single traps where the BCS to Bose-Einstein condensation transition 
takes place. These finite-size systems possess tunable interaction parameters which
can give rise to multiband superconductivity \cite{iskin} and are thus the subject of new promising experiments.

\acknowledgments
We would like to thank V.V. Moshchalkov for useful discussions.
VHD acknowledges the financial support by the grant EF/05/005 (INPAC) from the 
University of Leuven.


\begin{thebibliography}{0}
\bibitem{comment}
These are actually the gap functions $\Delta_{1,2} ({\mathbf
r})$ in Ref. \cite{dao}, respectively.
\bibitem{comment1} These phases have not been found in previous studies
\cite{peeters} because disks of smaller radii were considered.
\end{thebibliography}
\end{document}